\documentclass[10pt,twocolumn]{IEEEtran}
\usepackage{multicol}

\usepackage{color}
\usepackage{soul}
\usepackage{amsmath}
\usepackage{graphics}
\usepackage{setspace}
\usepackage{cite}
\usepackage{latexsym}
\usepackage{float}
\usepackage{epsfig}
\usepackage{multirow}
\usepackage[table,xcdraw]{xcolor}
\usepackage{cite,cases,url}
\usepackage{amssymb}
\usepackage{graphicx}
\usepackage{fancyhdr}
\usepackage{epstopdf}
\usepackage{balance}
\usepackage{subcaption}
\usepackage{enumerate}
\usepackage{array}
\usepackage{algorithmic}
\usepackage{algorithm}
\usepackage{enumitem}
\usepackage[normalem]{ulem}
\useunder{\uline}{\ul}{}

\newcolumntype{L}{>{\centering\arraybackslash}m{5cm}}
\newcolumntype{K}{>{\centering\arraybackslash}m{6cm}}
\newcolumntype{P}{>{\centering\arraybackslash}m{2.3cm}}
\newcolumntype{M}{>{\raggedright\arraybackslash}m{2cm}}
\newcolumntype{N}{>{\raggedright\arraybackslash}m{2.5cm}}

\fancypagestyle{firstpage}
{
    \fancyhead[L]{\footnotesize This article has been accepted for publication in the IEEE Network Magazine. Please cite it as A. S. Abdalla, P. S. Upadhyaya, V. K. Shah, and V. Marojevic, ''Toward Next Generation Open Radio Access Networks---What O-RAN Can and Cannot Do!,'' IEEE Network Magazine, May 2022. } 
    \fancyhead[R]{\footnotesize \thepage}
    \cfoot{}
    
}

\begin{document}

\title{%Next Generation Cellular Networks
Toward Next Generation Open Radio Access Networks---What O-RAN Can and Cannot Do!}
%\hl{\hl{Myths and Facts of} }
\author{
\IEEEauthorblockN{$^1$Aly S. Abdalla, $^2$Pratheek S. Upadhyaya, $^3$Vijay K. Shah and $^1$Vuk Marojevic \\
} 
\normalsize\IEEEauthorblockA{$^1$Dept. of Electrical and Computer Engineering,  Mississippi State University, USA \\
$^2$ Dept. of Electrical and Computer Engineering, Virginia Tech, USA \\
$^3$ Dept. of Cybersecurity Engineering, George Mason University, USA }\\
\normalsize\IEEEauthorblockA{Emails: asa298@msstate.edu, pratheek@vt.edu, vshah22@gmu.edu, vuk.marojevic@msstate.edu}
}
\maketitle
\thispagestyle{firstpage}
%TC:ignore 
\begin{abstract}
The open radio access network (O-RAN) describes an industry-driven open architecture and interfaces for building next generation RANs with artificial intelligence (AI) controllers. 
\textcolor{black}{We circulated a survey among researchers, developers, and practitioners to gather their perspectives on O-RAN as a framework for 6G wireless research and development (R\&D). The majority responded in favor of O-RAN and identified R\&D of interest to them. 
Motivated by these responses, this paper identifies the limitations of the current O-RAN specifications and the technologies for overcoming them. 
We recognize end-to-end security, deterministic latency, physical layer real-time control, and testing of AI-based RAN control applications as the critical features to enable and discuss R\&D opportunities for extending the architectural capabilities of O-RAN as a platform for 6G wireless.}

\end{abstract}
%TC:endignore
\IEEEpeerreviewmaketitle

\section{Introduction}
\label{sec:intro}

\textcolor{black}{Next generation} 5G and 6G networks are introducing architectural transformations from an inflexible and monolithic system to a flexible, agile and disaggregated architecture to support %cellular networks with
service heterogeneity, coordination among multiple technologies, and rapid on-demand deployments. %\cite{9247527}. 
Such transformations are enabled by the emerging open radio access network (O-RAN) framework which provides virtualization, intelligence, and flexibility while defining %. It is worth mentioning that the objective of O-RAN is to enhance the RAN performance through virtualization of network elements and 
open interfaces for network innovation. %~\sout{\cite{BONATI2020107516}}. % that are complemented with intelligence. 
% The distinguishing characteristic is the network intelligence that is %going to be 
% the pillar of O-RAN and 6G wireless networks for rapid deployment, operation, and optimization. 
%the provisioning of optimization,  and rapid deployment.
The O-RAN based 5G and future 6G networks will incorporate artificial intelligence (AI) into the deployment, operation, and maintenance of the network~\cite{bonati2021intelligence}. The O-RAN architecture is expecting to offer a new dawn of services and applications for cellular networks, such as virtual network slices and dynamic spectrum sharing.

O-RAN %captures a new trend in cellular network %of the future will be
%development, deployment, and operation. It 
takes advantage of Cloud RAN (C-RAN) principles and leverages the increasingly software-defined implementations of wireless communications and networking functions. 
Instead of legacy interfaces that are vendor-specific and controlled by major industry players, it defines open interfaces and an open architecture for innovation at all layers. 
It assumes that cellular network management will be increasingly  data driven %, employing machine learning (ML) techniques, 
and establishes the generic modules and interfaces for data collection, distribution, and processing.

%%\Vijay{Is this paragraph needed? We repeat most of it in II.A.} The O-RAN Alliance is the consortium responsible for maintaining the standard, i.e. architecture, components, processes, and interfaces, %i.e., O1, A1, and E2 deployed 
%%to enable disaggregation of radio access network (RAN) components. The O-RAN architecture introduces the central unit (CU), distributed unit (DU), and radio unit (RU), the interfaces among them and with the user equipment (UE), core network (CN), and the operations support system (OSS). This is illustrated in Fig.~\ref{fig:ORANStnds} and discussed in Section II. %The RU encapsulates the radio frequency and the low physical (PHY) layer processing blocks. The DU implements the higher PHY, medium access control (MAC) and radio link control (RLC) layers. The CU features the near-real-time RAN intelligent controller (near-RT RIC), where the artificial intelligence controllers  reside (Fig.~\ref{fig:ORANArch}).

There have been initial studies that analyze the O-RAN technology and possible use cases for improving the performance of cellular networks. %For example, the authors of
%Reference~\cite{niknam2020intelligent} demonstrates 
\textcolor{black}{The potentials of AI applied  on a %real-world 
%operator 
data set provided by a network operator %the O-RAN framework 
for radio resource management  are demonstrated in~\cite{niknam2020intelligent}}. A green O-RAN system model is proposed in~\cite{Pamuklu_2021}, which introduces the use of renewable energy sources and how to minimize the cost of deployment by applying reinforcement learning for dynamic function splitting.
%The work presented in
%\textcolor{blue}{The potentials and a few limitations of O-RAN are discussed} in~\cite{bonati2021intelligence} which provides 
A data-driven closed-loop control experiment with %deep RL
reinforcement learning agents for optimizing the scheduler of multiple RAN slices is presented in~\cite{bonati2021intelligence}. \textcolor{black}{A closed-loop control system for O-RAN is designed and validated on an outdoor testbed for maximizing the capacity of the system composed of an aerial and multiple terrestrial UEs by optimizing the drone location and transmission directionality~\cite{bertizzolo2021streaming}.} 

%\hl{Be explicit on the contribution w.r.t. related (magazine) papers.} 

\textcolor{black}{Prior work has presented the O-RAN architecture and use cases. The contribution of this paper is \textcolor{black}{to identify} the limitations of the current architecture and the technologies and opportunities for research and development (R\&D) for overcoming them. This is motivated by a survey that we circulated among %academic and industrial wireless 
researchers, developers, and practitioners to gather their perspectives on O-RAN as a framework %of interest 
for 6G wireless. % research and development (R\&D). 
Based on the responses that we received, we identify the critical components that are not addressed by the specifications---security, latency, real time control, and testing---and make recommendations.} %with emphasis on openness and transparency.

\textcolor{black}{The rest of the paper is organized as follows: %conducted a survey where we asked wireless experts about O-RAN or alternative 6G wireless frameworks and their research interest %as a motivation for this research 
%and present the results in Section II. 
Section II presents the survey results. We briefly introduce the O-RAN architecture, components, and interfaces in Section III, followed by the capabilities and use cases in Section IV. % from the perspective of network deployment and operation.
Section V identifies the main O-RAN limitations and directions for R\&D. Section VI provides the concluding remarks.}

\section{%Research
Community Survey}
\label{sec:survey}
%     \begin{figure*}
%   \centering
%   \begin{subfigure}[b]{0.32\textwidth}
%   \hspace{-64 pt}
%     \centering
%     \includegraphics[width=1.2\textwidth]{figures/ORAN/ORANsurveyPic1.eps}
%     %\caption{The reward}
%     \label{fig:ORAN1}
%   \end{subfigure}
%   \hspace{-62 pt}
%   \begin{subfigure}[b]{0.32\textwidth}
%   %\vspace{+30 pt}
%     \centering
%     \includegraphics[width=1.2\textwidth]{figures/ORAN/ORANsurveyPic2.eps}
%     %\caption{The distance}
%     \label{fig:ORAN2}
%   \end{subfigure}
%   \hspace{12 pt}
%   \begin{subfigure}[b]{0.32\textwidth}
%     \centering
%     \includegraphics[width=1.2\textwidth]{figures/ORAN/ORANsurveyPic3.eps}
%     %\caption{The power}
%     \label{fig:ORAN3}
%   \end{subfigure}
%   \caption{\hl{fonts too small in (b) and (c). Maybe put a, b left right and c below them on its own? Could we include more figures and any specific written answers that may be interesting to mention in the text?}. 
%   }%\hl{should the legend be DQN, or 'proposed DQN'? Is it easy to put DQN on top in legend?}}
%   \label{fig:ORANSurvey}
% \end{figure*}

\begin{figure*}[ht!]
\centering
%\hspace{-29 pt}
%\includegraphics[width=1.01\textwidth]{figures/ORAN/ORANPaperFlyer1.eps}\\
\includegraphics[width=1.01\textwidth]{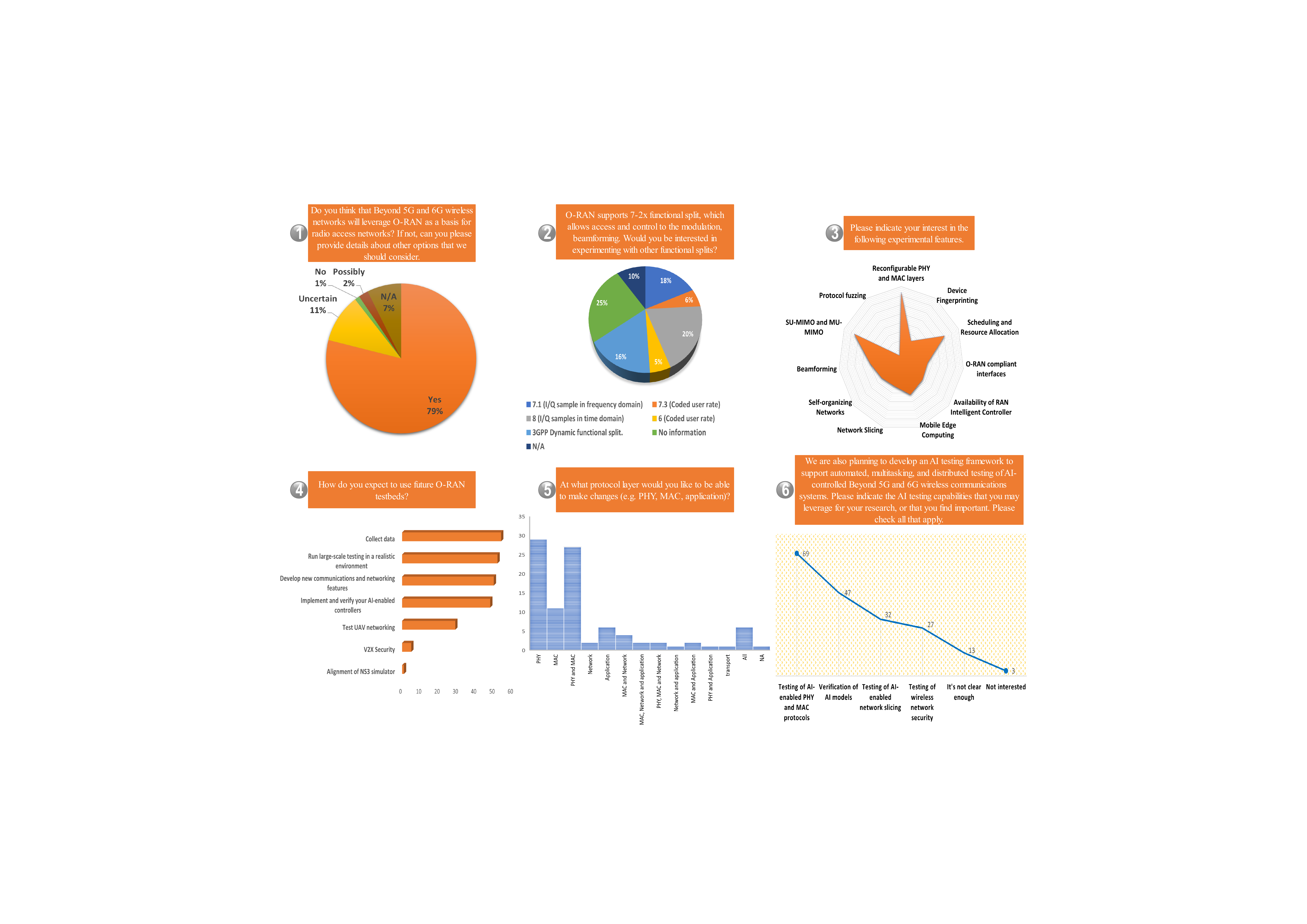}
\vspace{-2 pt}
\caption{Community survey outcome.}
\label{fig:ORANSurvey}
\vspace{-3mm}
\end{figure*}

In 2020/2021 we circulated a %Google
survey %~\cite{survey_link} 
among wireless researchers, developers, and practitioners to explore the interest in O-RAN %and future developments of O-RAN features %of the broader wireless %research community 
and to receive %their
feedback on research priorities. 
\textcolor{black}{About 150 experts in advanced wireless communications and networking were invited to participate in the survey. We were not seeking a popular poll and chose to distribute it via personal invitations to avoid misleading responses or misuse. Surveyors had to identify themselves and were asked for permission to use the anonymized data.} % whose affiliations are universities, research laboratories, and industry .

\textcolor{black}{Participants from $65$ institutions responded, $50$ located in North America, $12$ in Europe, and $3$ in Asia.} There were $95$ total participants: $21$ full professors, $13$ associate professors, $14$ assistant professors, $9$ postdoctoral researchers, $10$ \textcolor{black}{experienced PhD candidates}, and %the rest include 
$28$ experts from industry, government research laboratories, and research institutes. % from around the world. %Several academics from predominately undergraduate institutions participated.
%Out of the $95$ responses, 

%\textcolor{blue}{The purpose of this survey is to evaluate the interest in O-RAN and the specific features to explore in research.}
\textcolor{black}{The purpose of this survey was to initiate the dialog with the advanced wireless and networking R\&D community about O-RAN, understand if O-RAN is the framework of interest for 6G wireless R\&D, and what features are of most interest. We asked questions about what O-RAN capabilities they are most interested exploring and what new capabilities they would be interested in developing or using. For each question we provided a list of sample choices, allowed introducing alternative answers, and selecting multiple choices.}

Figure~\ref{fig:ORANSurvey}(1) shows that the majority agrees that O-RAN will be the foundation of %as a basic element for RANs in 
future cellular networks. %Seventy-nine percent of the survey takers believe in such scenario and only one percent of the survey takes disagree. %, while only $1\%$ who disagree and about one-tenth of the participants are not certain about that. 
The answers to Question 2 indicate that around 38$\%$ are interested in having the ability to access the in-phase and quadrature (I/Q) data in the time and frequency domains leveraging different functional split (FS) options. % along side with the 7-2x split option. 
Sixteen percent ($16\%$) %Also, there are 16$\%$ 
of the participants would like to see %encouraging 
support for a dynamic FS in % future 
O-RAN. % architecture as illustrated in Fig.~\ref{fig:ORANSurvey}(2).   
The O-RAN features of highest interest %to the wireless research community representatives 
are the programmable physical (PHY) and medium access control (MAC) layers, single and multi-user multiple-input, multiple-output (MIMO), scheduling and resource allocation %and mobile edge computing. This is shown in 
(Fig.~\ref{fig:ORANSurvey}(3)). %Moreover, there was a considerable appeal to elevate the current O-RAN attributes to support mobile edge computing.     
 %Some of the organizations who are interested in specific research topics have been summarized in Table~\ref{tab:T_FI}, extracted from the survey results. %(see attached supplemental document for full details of the survey). 
Figure~\ref{fig:ORANSurvey}(4) presents the expectations of the survey takers about their intended use of future O-RAN research testbeds. 
At the top of the list is data collection, large-scale experimentation under realistic constraints, development of novel communications and networking features, and the implementation and verification of AI-enabled wireless network controllers. Such use cases %and experiments are highly 
require access to %be performed at 
the PHY and MAC layers, as highlighted in Fig.~\ref{fig:ORANSurvey}(5).     
The majority of survey takers find it important to facilitate testing of AI controlled RANs, from the PHY to higher layers (Fig.~\ref{fig:ORANSurvey}(6)). %, most prominently  of AI-enabled PHY and MAC protocols.  
%%Overall, there is broad interest in %physical (PHY) and MAC functions and in 
%%using I/Q samples %for time and frequency functional splits 
%%for access %control to the 
%%to modulation, beamforming, and other PHY layer functions. %, access to MAC layer function and higher layer. 
%%Having a framework for collecting data and for large-scale testing of AI controlled RANs is of highest relevance. %~\cite{OAICURL}.

%%This is especially %Specifically, these new modifications will be critically 
%%required for AI-enhanced PHY and MAC layer functions, network slicing, and security algorithms according to Fig.~\ref{fig:ORANSurvey}(6).    
% desired by the survey respondents.
The obtained data motivates studying the capabilities and limitations of O-RAN, and what R\&D can do to extend the capabilities of the architecture and introduce complementary elements and processes.

%\section{From Legacy Cellular Networks  to %the 
%O-RAN} %Architecture}
\section{The O-RAN Architecture, Components, and Interfaces}
\label{sec:arch}
\begin{figure*}[ht]

\centering
\includegraphics[width=0.85\textwidth]{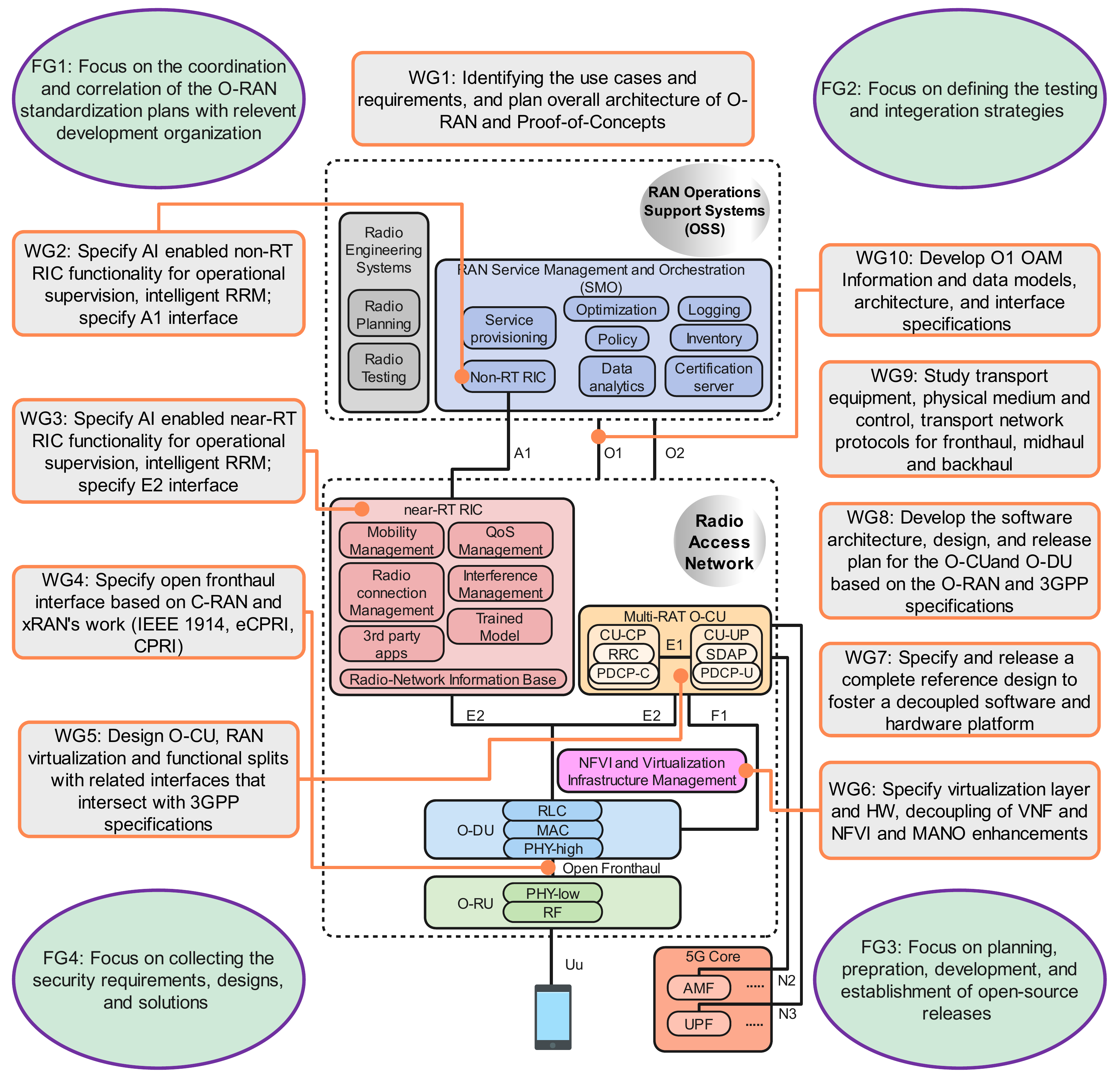}
\vspace{-2 pt}
\caption{The O-RAN architecture (center) and the work and focus groups with their objectives.} 
\vspace{-3mm}
\label{fig:ORANStnds}
\end{figure*}
The O-RAN Alliance was formed by merging the C-RAN Alliance %~\cite{} 
and the xRAN Forum. %~\cite{}
Its %leads the open RAN movement that 
mission is to extend the current RAN standards and facilitate open, intelligent, virtualized, and fully interoperable next generation RANs~\cite{ORANwhitepaper}. 
Many of its members are also members of \textcolor{black}{the} Third Generation Partnership Project (3GPP). 
%%, the O-RAN Alliance extends the 3GPP standards %to bring the industry together 
%%with new interfaces and intelligent controllers, creating a framework for developing and deploying intelligent, software-defined networking (SDN) based, and virtualized cellular networks.
%The O-RAN Alliance is thus a consortium that develops standards for open cellular networks.
They meet regularly to update specifications that are %published and 
made accessible to the community. % every three months. 
The O-RAN Alliance
defines study items %the cover the overall O-RAN architecture and is 
that are currently organized %divided 
into \textcolor{black}{ten} technical work groups (WGs) and four focus groups (FGs).
Figure~\ref{fig:ORANStnds} shows the high-level O-RAN architecture and the roles of the WGs and FGs. %, and the timeline of the O-RAN software releases. 

%\subsection{The O-RAN Architecture}
The O-RAN architecture %as depicted in the center of
%Fig.~\ref{fig:ORANStnds} is based on the reference architecture introduced in~\cite{ORANwhitepaper}. 
%It 
encompasses the RAN, %the main
%components of 
the operations support systems (OSS), %\sout{the radio engineering systems, }
and %at the top that are %deployed for 
%performing the orchestration and management of RAN functions through 
open interfaces. 
%%\begin{figure}
%%\label{fig:results3}
%  \subfloat[]{
%	\begin{minipage}[c][0.55\width]{
%	   0.55\textwidth}
%	   \centering
%	   \includegraphics[width=0.7\textwidth]{Figures/Secrecy_2UEs_UAV_TO_Iterations_TextUpdate.eps}
%	    \label{fig:results_ite2}
%	\end{minipage}}
% \hfill 	
%  \subfloat[]
%  {
%	\begin{minipage}[c][0.55\width]{
%	   0.4\textwidth}
%%	   \centering
	   %%\includegraphics[width=0.4\textwidth]{figures/ORAN/ORAN_GeneralArchitecture1.pdf}
%	\end{minipage}}
%%	\vspace{-2 pt}
%%\caption{O-RAN general architecture. \hl{Seems like the DU is connected to both near-real time RIC and CU through E2 interface-is that right?} %(a) two UEs, and (b) 
%%\label{fig:ORANArch}
%%}
%%\vspace{-3mm}
%%\end{figure}
The design of the O-RAN architecture is in harmony with the %long term evolution (LTE) and 
5G new radio (NR) RAN architecture defined by %the 
3GPP. The central unit (CU), distributed unit (DU), and radio unit (RU) of a modern RAN %within the O-RAN architecture 
are responsible for delivering diverse functions of the radio protocol stack associated with the radio resource control (RRC), \textcolor{black}{service data adaption protocol (SDAP), packet data convergence protocol (PDCP)}, radio link control (RLC), MAC, and PHY layers. %These are the O-RAN CU (O-CU), O-DU, and O-RU in the O-RAN architecture.
%from the old versioned RAN.
The O-RAN RU (O-RU) is in charge of %the attachment of 
establishing the PHY layer connection with the user equipment (UE). %via the Uu interface. 
It %and/or the 
integrates the antenna elements with the radio frequency (RF) processing components, such as transceivers, analog beamformers, and power amplifiers. In addition to the RF processing, the O-RU integrates the lower level PHY layer processing (PHY-low), such as digital beamforming and the fast Fourier transform (FFT)/inverse-FFT. %, will be performed in the . 
The Open Fronthaul interface %is utilized to
connects the O-RU to the O-DU. 
The O-DU is responsible for the higher level processing functions at the PHY layer (PHY-high), such as channel modulation and coding/decoding, the MAC layer, %such as data transfer and radio resource allocation, 
and the RLC layer. % for error correction and segmentation, among others. 

The upper layers of the radio protocol---the RRC, the PDCP, and the SDAP---are provided by the O-CU. The design of the O-CU %within the O-RAN architecture 
enables the simultaneous operation of both long-term evolution (LTE) O-DUs and 5G NR O-DUs, and is known as multi-radio access technology (multi-RAT). It is supported by the network function virtualization infrastructure (NFVI).
The N2 and N3 interfaces %of The 3GPP %for 5G NR 
are responsible for handling the communications between the multi-RAT O-CU %of O-RAN 
and the access \textcolor{black}{and} mobility management function (AMF) and user plane function (UPF) %, encapsulated in the ..... (AMF) and .... (UPF) 
of the 5G core network. 

%\sout{Whereas the O-DU and O-RU rely on specific hardware implementation}, \hl{Is this (before and after) true?} the network function virtualization infrastructure (NFVI) is deployed within the O-RAN architecture to enable the execution of the multi-RAT O-CU. Additionally, to support the disaggregation strategy, 
3GPP introduced the F1 interface that connects the CU to the DUs and the E1 interface to enable coordination between the control and user planes at the CU. % units. 
The O-RAN Alliance %parallelly
establishes the E2 interface, which %enables the control and optimization of the protocol stack functions of the RAN through the near-RT RIC. %, and the A1 interface, which communicates with the SMO. 
%More precisely, the E2 interface %is responsible for 
forwards the %collected %\sout{RAN} 
measurements from the O-DU \textcolor{black}{and O-CU} to the near-RT RIC, and the %needed
configuration commands back to the O-CU and O-DU. \textcolor{black}{The near-RT RIC is a logical function that enables near-RT RAN control and optimization. This is implemented through %third-party 
%applications called 
xApps.} 

The OSS is a component of modern and increasingly software-defined wireless networks. %~\cite{OSS}. %Its role in O-RAN will be discussed here. 
It is responsible for radio planning and testing and for service management and orchestration (SMO), i.e. for monitoring, and life cycle operation and management functions of the softwarized O-RU, O-DU, O-CU, and near-RT RIC components. % of the O-RAN. % in the O-RAN architecture. 
There are different subsystems within the OSS for, among other things, %such as service provisioning, %optimization, policy, 
data analysis, logging, inventory, certification, and non-RT RIC operations \textcolor{black}{through third-party rApps}. %%These %various
%%subsystems %included in the OSS 
%%are in charge of representing the services and network components, designing and executing different optimizations, creating and operating policies to manage other OSS functions, and collecting and storing data streams for analysis. The service management and orchestration (SMO) is a major component of the OSS. % and performs three main functions. 
%%It %The SMO framework 
%%manages the fault, configuration, accounting, performance, and security %(FCAPS) 
%%of O-RAN, optimization of the RAN, and orchestration and management of O-RAN cloud platforms. 

The A1, O1 and \textcolor{black}{O2} interfaces %are implemented to
assist with the signaling between the OSS and the RAN.  %underlying O-RU, O-DU, and O-CU nodes. 
The A1 interfaces facilitates AI related parameter exchange between the non-RT and near-RT RICs. The O1 interface, which is standardized by 
3GPP, is used for fault, configuration, and performance management of the RAN nodes. The \textcolor{black}{O2} interface facilitates resource management at the NFVI of virtualized RAN deployments. %at the NEVI. 

%%The non-RT RIC is a controller %integrated within the OSS platform for 
%%that provides policy management, AI management, and data enrichment services to the underlying RAN nodes. The AI management service %of the non-RT RIC 
%%includes the transfer of the trained model to the near-RT RIC and the periodic or triggered updates of the AI model. The data enrichment capability of the non-RT RIC %focusing on
%%delivers the required information to the near-RT RIC, which performs data analytic and management functions\sout{ to draw conclusions, make predictions, and drive informed decisions}. %The interactions between the near-RT RIC and non-RT RIC happen over the A1 interface. In addition to the A1 interface, the O1 and \textcolor{blue}{O2} interfaces %are implemented to
%assist with the signaling between the OSS and the underlying O-RU, O-DU, and O-CU nodes: 

%\textcolor{red}{Paragraph on ORAN alliance, WG and timeline.}
%\textcolor{blue}{Elaborate on the planning, testing, operations, and the specific interfaces. Also mention if open MANO, or other OSS implementations, open source or from traditional vendors, like Ericsson, can be applied to manage O-RAN. 
%\\
%Should we refer to Fig. 2 in this section somewhere, to break the text?}

%\textcolor{red}{Could we compare 3GPP interfaces (CPRI, eCPRI) vs. O-RAN)? Illustrate/examples vendor specific vs. open. Could be in the introduction section.}

%\section{O-RAN Resource}
%\label{sec:resources}
%\input{include/resources.tex}

\section{%What 
\textcolor{black}{What O-RAN Can Do: Capabilities and Use Cases 
}}
\label{sec:services}

\textcolor{black}{O-RAN is disruptive in that it opens up the RAN interfaces to spur innovation and contributions for new developers and vendors. %, virtualizes the RAN software, and several other capabilities and usecases detailed in this section.
This can provide several benefits, such as % to mobile network operators -- 
reduced cost of maintenance, %multi-vendor ecosystem, 
dynamic services, and quicker time to market for new user and network management services as well as other innovations in the wireless domain. This section summarizes the salient features of O-RAN under three categories.}

%\textcolor{blue}{Features: disaggregation-standard interfaces-multi-vendor support, intelligence, different timescales that are supported, openness, software, \\

%Introduce those, if not already, and explain what these features enable...}

%\input{Tables/TechnologiesTable}
\begin{table*}[ht]
\centering
\caption{Representative O-RAN use cases employing AI \textcolor{black}{as defined by the O-RAN Alliance~\cite{li2020ran}}. %\hl{Add a few xApp use case that have demonstrated O-RAN.}
}
\footnotesize
%\resizebox{0.85\textwidth}{!}{%
%\begin{tabular}{|c|c|c|c|}
{\begin{tabular}{|p{1.7cm}|p{0.9cm}|p{3.5cm}|p{0.9cm}|p{1cm}|p{4.4cm}|p{2.7cm}|}
\hline
 \textbf{Use case} &\textbf{Network layer} &  \textbf{AI %models
 functionality description}  &   \textbf{Training host}  & \textbf{Inference host }  &   \textbf{Training data}  &  \textbf{Action} 

\\ \hline
%\vspace{0.001 in} 
Massive MIMO beamforming optimization 
& PHY  
& Predict the optimal configuration of massive MIMO parameters for each cell according to a global optimization objective designed by the operator. 
&  Non-RT RIC &  Non-RT RIC 
& Key performance indicators (KPIs) related to traffic load, coverage and QoS performance, per beam/area and massive MIMO configuration.
& Configure the optimized beam parameters via the O1 interface.
\\ \hline

%\vspace{0.001 in}
Quality of experience (QoE) optimization 
& MAC  
& %Making predictions on
Predict the application/traffic types, QoE and available bandwidth. 
& Non-RT RIC 
& Near-RT RIC 
& Network level measurement report including UE data, L2 measurement, RAN protocol stack status, cell level information, QoE measurement metrics and user traffic. 
& E2 control or policy commands towards the RAN for the QoE optimization.
\\ \hline

%\vspace{0.001 in} 
Dynamic spectrum sharing (DSS) 
& MAC  
& Predict the short-term traffic demand %for 4G and 5G network 
based on near-RT metrics from the RAN. 
& Non-RT RIC  
& Near-RT RIC 
& Location, DSS modality, cell configuration, used, reserved, requested, and blocked PHY resource blocks (PRBs), number of active UEs, traffic demand, QoS classes, and UE capability %and UE location 
information. 
& E2 control/policy of resource configuration and scheduling coordination.
\\ \hline

%\vspace{0.001 in} 
Network slice subnet instance (NSSI) resource allocation optimization 
& MAC  
& Predict of the traffic demand patterns of NSSI at different times and locations. 
& Non-RT RIC & NA 
& Measurement counters per slicing subnet instance. These  include downlink (DL) and uplink (UL) PRBs used for data traffic, average DL and UL throughput, % in gNB average UL UE throughput in gNB, 
and the number of %requested
packet data unit sessions % to setup, Number of PDU sessions 
successfully set up and %number of PDU sessions 
failed to set up.
& Reconfigure the NSSI attributes via the O1 interface and update the Cloud resources via the O2 interface.
\\ \hline
Context-based dynamic handover (HO) management for vehicular users 
& MAC  
& Customize HO sequences for a UE according to intents, policies, and configurations taking into consideration the high speeds and the heterogeneous environment of vehicular UEs.
& Non-RT RIC 
& Near-RT RIC 
& Historic data from %\hl{V2X} 
a vehicular UE over the V1 interface, which include %V2X 
UE location, velocity, and trajectory, %, navigation information. %, \hl{GPS} data. %, CAMs, DENMs. 
measurement reports for serving and neighboring cells, and connection and mobility statistics. 
& RAN deploys the configuration received from the near-RT RIC over the E2 interface.
\\ \hline
%Flight path based dynamic 
Radio resource allocation to unmanned aerial vehicles (UAVs)
& MAC  
& Perform the radio resource allocation %for on-demand coverage 
for UAVs to tackle %introduced 
challenges %as the height increase, 
such as %\hl{LoS} propagation/UL 
RF interference, poor KPIs caused by base station antenna patters (side lobes towards sky). %, and sudden drops in signal strengths.  
& Non-RT RIC 
& Near-RT RIC 
& Aerial vehicle related measurement metrics %from network 
from SMO and based on UE measurement reports, radio channel information, mobility related metrics, flight path, climate, %flight
forbidden/limited 3D area %and space load 
information, etc. %from application, e.g. unmanned aircraft system traffic management. % (UTM). 
& The near-RT RIC delivers the radio resource allocation configurations to the O-CU/O-DU over the E2 interface.
\\ \hline
Traffic steering
& MAC  
& Traffic steering among various RATs, such as LTE, NR, and Wi-Fi, %to enable balanced distribution of traffic 
according to different traffic management    objectives. 
& Non-RT RIC 
& Near-RT RIC 
& Measurement reports, cell load statistics % including intra-RAT and inter-RAT, 
cell quality thresholds, %\hl{CGI}
channel quality indicator reports and measurement gaps on per-UE or per-frequency basis. %Also, cell load statistics including number of active users or connections, number of scheduled users per TTI, and PRB utilization. %, and \hl{CCE} utilization. 
& The traffic management %optimized
policy is activated and transferred to the %E2 nodes 
near-RT RIC via the A1 interface. % to be enforced.
\\ \hline
\end{tabular}%
}

\label{tab:ORANuse}
\end{table*}

%%\subsection{Dynamic Disaggregation %, Functional Splits, 
%%and Multi-Vendor Support %\hl{combine with C?--there is a lot of repetition and may be difficult to split cleanly}
%%}

\subsection{\textcolor{black}{RAN Disaggregation, Open Interfaces, and Multi-Vendor Support}}

\begin{figure}[t]
	   \centering
	   \includegraphics[width=0.49\textwidth]{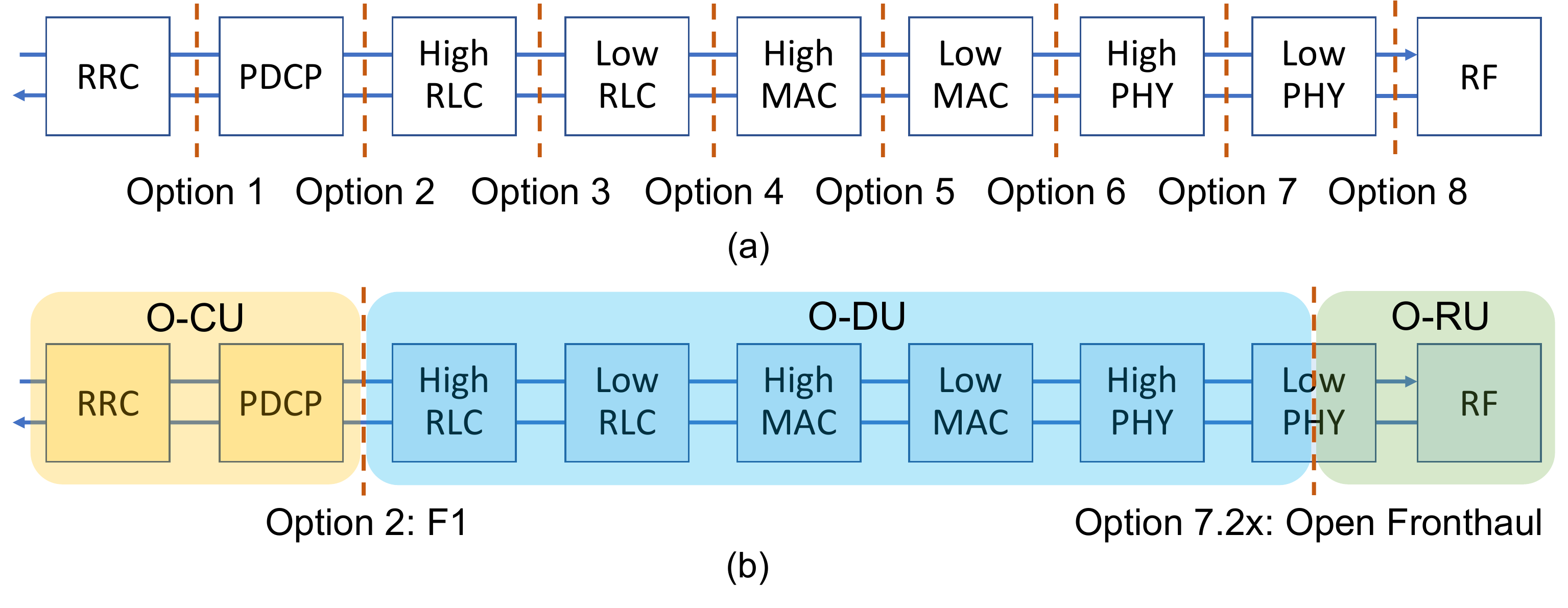}
\vspace{-6mm}
\caption{3GPP FS options (a) and the O-RAN FSs and interfaces between the O-CU, O-DU, and O-RU (b).}
\label{fig:fs}
\end{figure}

\textcolor{black}{Functional splits were introduced in 3GPP Release~15 to allow splitting the base station 
functionalities into the CU and the DU, implementing the higher and lower layers of the RAN protocol stack, respectively. 
Although %\hl{The} 
3GPP defines many split options (Fig.~\ref{fig:fs}a), % different levels, 
{vendors} use %\textcolor{blue}{closed standards}
\textcolor{black}{proprietary implementations and interfaces}
%\sout{lock-in} 
which has led to single vendor network solutions. %in practice implies that the software which implements the CU and the DU is limited to the chosen vendor. 
O-RAN adopts %the 3GPP 
FS Option 2 for the F1 interface between the O-DU and O-CU%(higher-layer split--HLS)
, and %\hl{the O-RAN specific} 
Option 7-2x for implementing the fronthaul, the interface between the O-DU and O-RU% (lower-layer split--LLS)
%, because it provides the best support for 4G/5G systems
~(Fig.~\ref{fig:fs}b). %~\cite{REF?}. 
%The CPRI is based on the lower layer split 8, where the RF unit is detached from the baseband processing. The needs for more flexibility, reliability with higher transmission rates have motivated the O-RAN to employ a modified version of the CPRI, which is known as the enhanced CPRI (eCPRI) for the fronthaul specifications. 
O-RAN employs a modified version of the common public radio interface (CPRI), the enhanced CPRI (eCPRI), for the fronthaul~\cite{ORANFH}.}

\textcolor{black}{The openness of the O-RAN architecture and interfaces enables cellular operators to employ unique strengths of each vendor for a delicate application or service and facilitates RAN function sharing. % purposes. 
It is envisaged that the near-RT RIC and non-RT RIC will be able to support different vendors' O-CUs and O-DUs. A direct benefit of the disaggregated RIC architecture and open interfaces (E2 and A1) is that it supports interoperability and flexibility in deployment and independent fault tolerant systems. An example of this is the collaboration between the O-RAN Software Community (OSC) and Open Networking Foundation's SD-RAN %\cite{}
and the compatibility between OSC's RIC and software radio solutions such as srsRAN and OpenAirInterface \cite{johnson2022nexran,polese2021colo}.}%to efficiently use the resources for different use cases. 
%%Unlike traditional RAN architectures, O-RAN envisions the use of standardized interfaces to make network sharing practical, %a key feature 
%for reducing cost and s known as . 
%Here, due to the openness and virtualization of RAN components
%%where physical infrastructure operators can %share their wireless infrastructure and 
%%host %logical 
%%CU and DU implementations as virtual network function (VNFs) from multiple vendors~\cite{5GWF18}.

\textcolor{black}{Depending on the bandwidth and latency requirements, it is possible to deploy the \textcolor{black}{O-CU, O-DU, and near-RT RIC functions} %either 
centrally or at the edge. %, close to the O-RUs. 
%O-CU and O-DU functions 
They can be implemented as virtual network %software
functions (VNFs) and be deployed in any RAN tier on commercial off-the-shelf hardware, as has been demonstrated with software radio implementations of 4G/5G RANs~\cite{AlyCommMag22}. 
In practice, these %splits 
are implemented as virtual machines and containers managed by a hypervisor}\footnote{\textcolor{black}{A hypervisor, also known as virtual machine monitor, is software that creates and controls virtual machines.}} \textcolor{black}{and virtualized infrastructure manager~\cite{5GWF18}, 
%The hypervisor allocates and manages virtual machines. The VIM deals with the allocation of resources in the NFV infrastructure (NFVI). They include computational resources (processors), storage, and network resources.
respectively, to ensure scalability, diverse network support, and interoperability. 
Load management, 
RT performance optimization, and maintenance of various quality of service (QoS) requirements, %for different applications %like gaming and voice
among others, can thus 
be handled through multiple instances of the O-CU and the O-DU that share the same physical resources.} %, thereby reducing cost and increasing network performance. 

\subsection{Support for Different Timescales} %Non and Near-Real-Time %Timescale 
%Support}
%\textcolor{blue}{This subsection can come before the previous. I suggest introducing the 3 AI control loops in this section, unless they fit better in Section II, for instance.} %\textcolor{blue}{Do we need a subsection or could it be a simple paragraph under Integration of AI/ML?}
% \Vijay{Also, this time scale is not just about AI.. It's about RAN control loops like, it is okay to do slicing in minutes or hours, whereas scheduling needs to be done in milliseconds. Beamforming may be required to be done in milliseconds or seconds depending upon the mobility of users. See the control loops table in Northeastern paper.}
One of the main requirements for meeting the heterogeneous 5G services\textcolor{black}{, such as interference management, resource allocation, security, and traffic offloading,} %with the O-RAN architecture 
is being able to perform RAN operations at the appropriate timescales. A 5G network may respond with resource changes %on the 
at sub-millisecond time granularity. %order. % of $1$ ms. %~\cite{5G_book}. %, whereas 6G systems are expected to operate respond to an order of magnitude faster. To enable this, 
The O-RAN Alliance defines the closed-loop control of the RAN by enabling the RICs to operate at different timescales: 
\begin{itemize}
    \item \textbf{The non-RT control loop} %- The control loop 
    operates at a timescale of at least 1 s through the A1 and O1 interfaces. Examples include instantiation and orchestration of network slices, resource allocation at the infrastructure level and data driven RAN policy guidance to xApps in the near-RT RIC. Such decisions can be taken at frequencies spanning seconds, hours, or even days.
    \item \textbf{The near-RT control loop} %- This control loop 
    operates on a timescale between 10 ms and 1 s through the E2 interface. The xApps %hosted on the near-RT RIC %serve \hl{lighter detection purposes} and
    leverage user session data and MAC and PHY layer data to optimize the %user 
    quality of experience (QoE) 
    by controlling time sensitive services, such as resource scheduling, beamforming, load balancing, and handover. % processes. 
    % \item real time control loop - This control loop takes care of the operations of a cellular network at the sub-10 ms or sub-ms timescale. These sub-ms tasks are currently not a part of O-RAN implementation since it requires high level computation and RU level standardization.
\end{itemize}
%This will require AI output with very low latency.
Combining the fast processing at the near-RT RIC with the larger timescale analysis at the non-RT RIC %and SMO 
is a unique capability of the O-RAN architecture. \textcolor{black}{Table~\ref{tab:ORANuse} shows some of the RAN operations that can be implemented leveraging the different timescale properties of O-RAN enabled by the non-RT and near-RT control loops.}% that aims to provide O-RAN with a small response time. 

\subsection{\textcolor{black}{AI Integration and xApps/rApps}}

The O-RAN architecture %design 
%provides the capability of 
inherently enables %simultaneously 
collecting %the %required 
application %information 
and radio environmental information to perform intelligent decisions %based on the retrieved information 
using AI models. %Such capability is not supported by the \hl{conventional and emerging}\textcolor{red}{``conventional and emerging'' needs to be changed or explained as per reviewer 1.10 comment} base stations, but is supported by design of the O-RAN specifications. 
%\textcolor{blue}{%Whereas recent advances in intelligent automation of 3GPP networks are provided by the network data analytics function (NWDAF), such automation is focused % are limited to 
%on %supporting 
%the
%5G core network.
% Whereas 3GPP introduces the network data analytics function for standardizing data collection, production, and use in the 5G core network, %intelligent automation of 3GPP networks are provided by the network data analytics function (NWDAF), such automation is focused % are limited to 
% %on %supporting 
% %the
% %5G core network and 
% it does not offer open standard application programming interfaces.} %gNB/eNB of 5G and 4G cellular networks. 
\textcolor{black}{The non-RT RIC is a controller of the SMO %platform for 
that 
provides policy management, AI management, and data enrichment services to the underlying RAN nodes. AI management services %of the non-RT RIC 
include the transfer of the trained AI model to the near-RT RIC and the periodic or triggered updates of the model. The data enrichment capability of the non-RT RIC %focusing on
delivers the required information to the near-RT RIC, which performs data analytics and management functions.}

The variability of key performance indicators (KPIs) and QoS levels of %the provided 
5G use cases %and applications 
along with the existing operation and maintenance functions of cellular networks enable %that are able only on 
optimizing the cellular performance parameters per cell. % not individually.
The O-RAN technology, on the other hand, enables performance enhancements per UE from the collected long-term traffic, coverage, and observed interference information, among others. %, an then performing it via the near-RT RIC to the gNB for execution.
Integrating the resource management %scheduling process of different 
 %from the core network %'s QoS mechanism 
in the near-RT RIC %function module 
is beneficial for reducing the burden on the core network as well as reducing the data transfer and time overhead of forwarding and processing data between the RAN and the core network. This %e newly introduced intelligence at the RAN
%can  time in transmission process, which overall 
improves the overall efficiency and latency of the system.

% \begin{itemize}
%      \item \textbf{QoE Optimization}: this use case automates the configuration and controlling of RAN parameters to improve the QoE of the user by utilizing the intelligent and open interface of the O-RAN architecture. The limited radio resources are utilized in a way that the  UE's use specific applications that requires critical KPI e.g. Cloud VR assign sufficing radio resources to meet these KPI and improve there QoE. The ML solution for this case consists of training model at the Non-RT RIC to obtain QoE related models such as Application Classification Model, QoE Prediction Model, QoE policy Model and Available BW Prediction Model. The offlkine model will use the training data data in table~\ref{tab:ORANuse} which provided by the E2 node over the O1 interface. After the training phase, the QoE related ML model will be deployed in Near- RT RIC to assist in the QoE Optimization function. The Near- RT RIC
%  \end{itemize}

%\textcolor{blue}{The near-RT RIC can host third-party control applications, i.e. xApps.} 
AI is expected to improve several functionalities of software-defined RANs. Among those functionalities are load balancing, resource allocation, fault detection, and security. %%Table~\ref{tab:ORANuse} provides details about \textcolor{blue}{various} use cases \textcolor{blue}{for different applications and network functions }that are defined by the O-RAN Alliance~\textcolor{blue}{\cite{li2020ran}}. It describes several opportunities for AI \textcolor{blue}{as xApps in the near-RT RIC or rApps in the non-RT RIC} for practical cellular networks implementing the O-RAN interfaces. %SDN controllers have the greatest level of access to the system making them a high value target to attackers. 
\textcolor{black}{Table~\ref{tab:ORANuse} summarizes the currently defined use cases~\cite{li2020ran} where the O-RAN architecture can provide a major impact in terms of AI control and enhanced RAN performance. %It details the entire AI control loop for the use cases and the main requirements to achieve this. 
It identifies the layers at which the network function is implemented, the AI operations, and the type of data that is required to train these AI models. % has been noted. 
Preliminary works implementing these use cases with open source software include 
\begin{itemize}
    \item An xApp implementing a basic network slicing framework in srsRAN \cite{johnson2022nexran},
    \item A slice and resource scheduling xApp employing deep reinforcement learning with a fully functional data collection framework \cite{polese2021colo}, and
    \item A traffic steering xApp developed by the OSC to showcase dynamic handover of users from one cell to another based on traffic predictions \cite{dryjanski2021toward}.
\end{itemize}
} 

\textcolor{black}{Furthermore, besides collecting and training AI models for xApps in the near-RT RIC, the non-RT RIC can itself host AI-driven %third-party control applications (
rApps, such as AI-based orchestration of network slices and data-driven policy management, to improve the automation and management of RAN resources and %near-RT RIC 
xApps.}

\section{{\textcolor{black}{What O-RAN Cannot Do: Limitations and R\&D Directions}} %What O-RAN %Can and 
%Cannot Do%\hl{---Limited Capabilities/Features}
}
\label{sec:capabilities}
\textcolor{black}{This section identifies %and provides insights into 
the major limitations of the current O-RAN specifications and outlines R\&D directions to overcome those and spur the evolution of O-RAN %a wide adoption of O-RAN 
as the architectural framework for next generation wireless network deployments.}

%\textcolor{blue}{For selected topics of relevance, discuss the capabilities of ORAN and the non-capabilities and those may be enabled}

%The O-RAN architecture and processes have certain restrictions and lack a few important capabilities/features that we discuss here.
   
\subsection{End-to-End Security} %Issues and Vulnerabilities}
    The O-RAN architecture enables the dynamic disaggregation of functionalities, the introduction of new functions, protocols, components, and interfaces. The new components, interfaces, and the lower layer split (LLS) however expands the threat surface and makes O-RAN %more
    prone to additional security risks beyond those of the 3GPP architecture. \textcolor{black}{The O-RAN security threats can be categorized in six primary classes~\cite{SFG}}:
    \begin{enumerate}
        \item The first threat class is related to the new functions that O-RAN adds, which include the SMO \textcolor{black}{and} the %non-RT RIC, and Near-RT 
    RICs. A misconfiguration or missing authentication process related to these functions can become a severe threat to network operations. % for the O-RAN. 
        \item The second class is characterized by the improper or missing ciphering of the data sent across the open interfaces A1, E2, O1, O2, and the fronthaul. % enables effective penetration attacks.
        \item The third class of attacks comes from the LLS%modified architecture
    ---the 7-2x split---that introduces vulnerabilities via possible decision conflicts among the O-RAN components. 
        \item The \textcolor{black}{fourth} threat class results from the functional decoupling and multi-vendor support. This can lead to the reliance on different security levels and the lack of a %with unavailable 
    root of trust. %via various vendors increases the penetration rate of the open interfaces via heterogeneous security levels. 
        \item The fifth class comes from the disaggregation of software and hardware %via mixed containerization 
    and the virtualization of the O-RAN system that lacks sufficient security measures.
    %makes the network more vulnerable to reverse attacks.
        \item The sixth class stems from the implementation of O-RAN using open-source software. This raises security risks where an adversary may be able to %analyze and  
    replicate and test the software and system operations to find loopholes %that can be compromised by enabling backdoors 
    and design %launch 
    attacks that may be difficult to detect.  
    %The disaggregation of hardware and software, virtualization, and use of open source components also present security risks that must be addressed. 
    
    \end{enumerate}

\begin{figure}[t]
\centering
\includegraphics[width=0.49\textwidth]{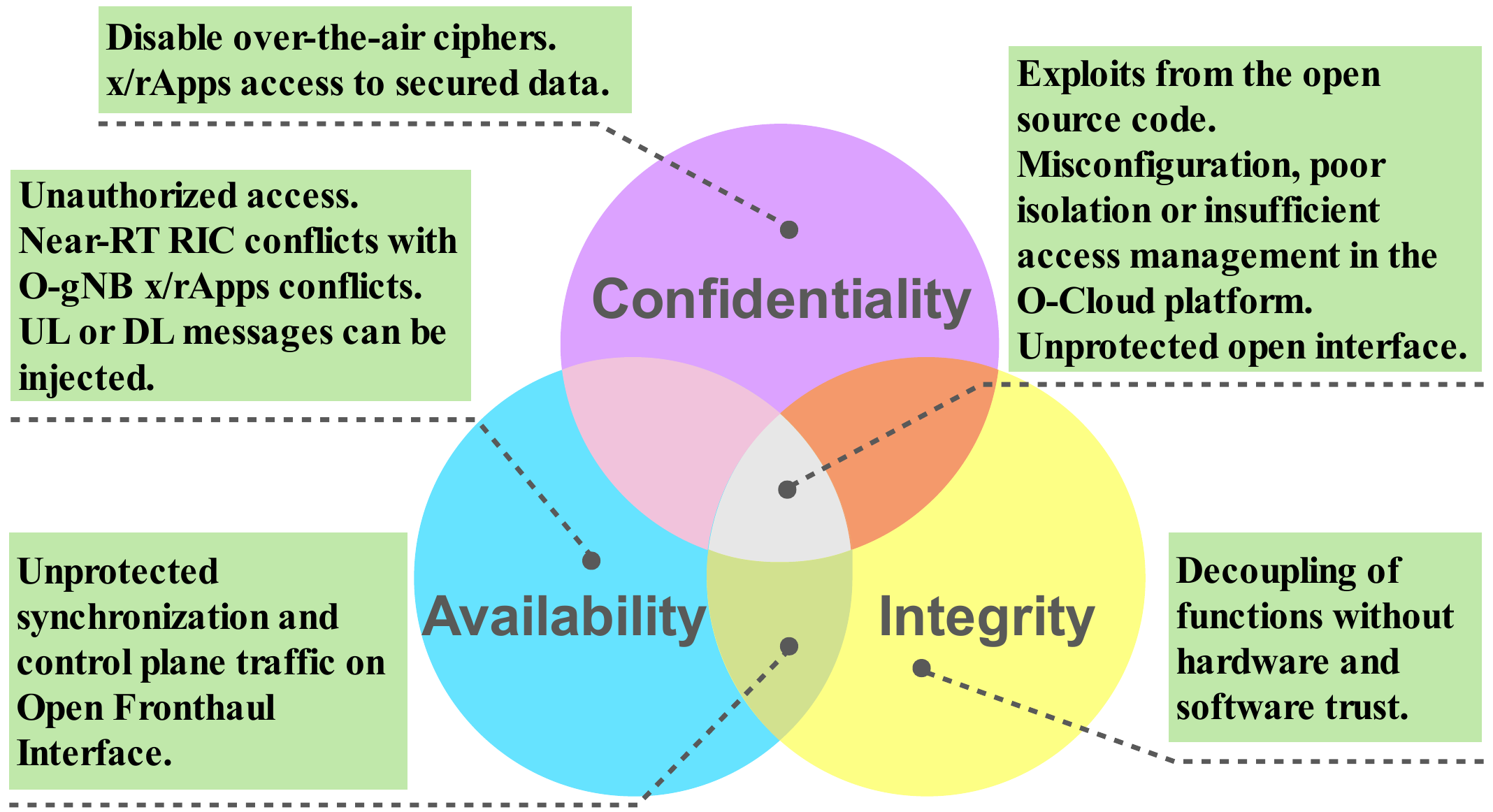}
\vspace{-3mm}
\caption{Potential security vulnerabilities of O-RAN.}
\label{fig:ORANSec}
\end{figure}

    Figure~\ref{fig:ORANSec} illustrates the potential security vulnerabilities that could be exploited to compromise %attacks against
    the confidentiality, integrity, and availability of next generation wireless networks implementing the O-RAN architecture. Security threats can target the %O-RAN
    architecture, %the O-RAN Cloud,
    the open-source software-defined procedures, and the AI subsystem. 
    An example of a security attack would be hijacking an operator's RAN infrastructure and resources by exploiting the vulnerabilities of remote access privileges provided to other operators in the RAN sharing scenario. 
    \textcolor{black}{Hijacking is a type of network security attack in which the attacker takes control of one or more %core 
    network functions, the open-source software, %and 
    or the %implemented 
    AI subsystem in this case. Exploiting a vulnerability refers to a set of actions that an attacker takes after finding a weakness in the hardware or software to proceed with a targeted attack.} %Proper measures need to be taken to secure communication/control channels }
    
\begin{figure}[ht]
\centering
\includegraphics[width=0.49\textwidth]{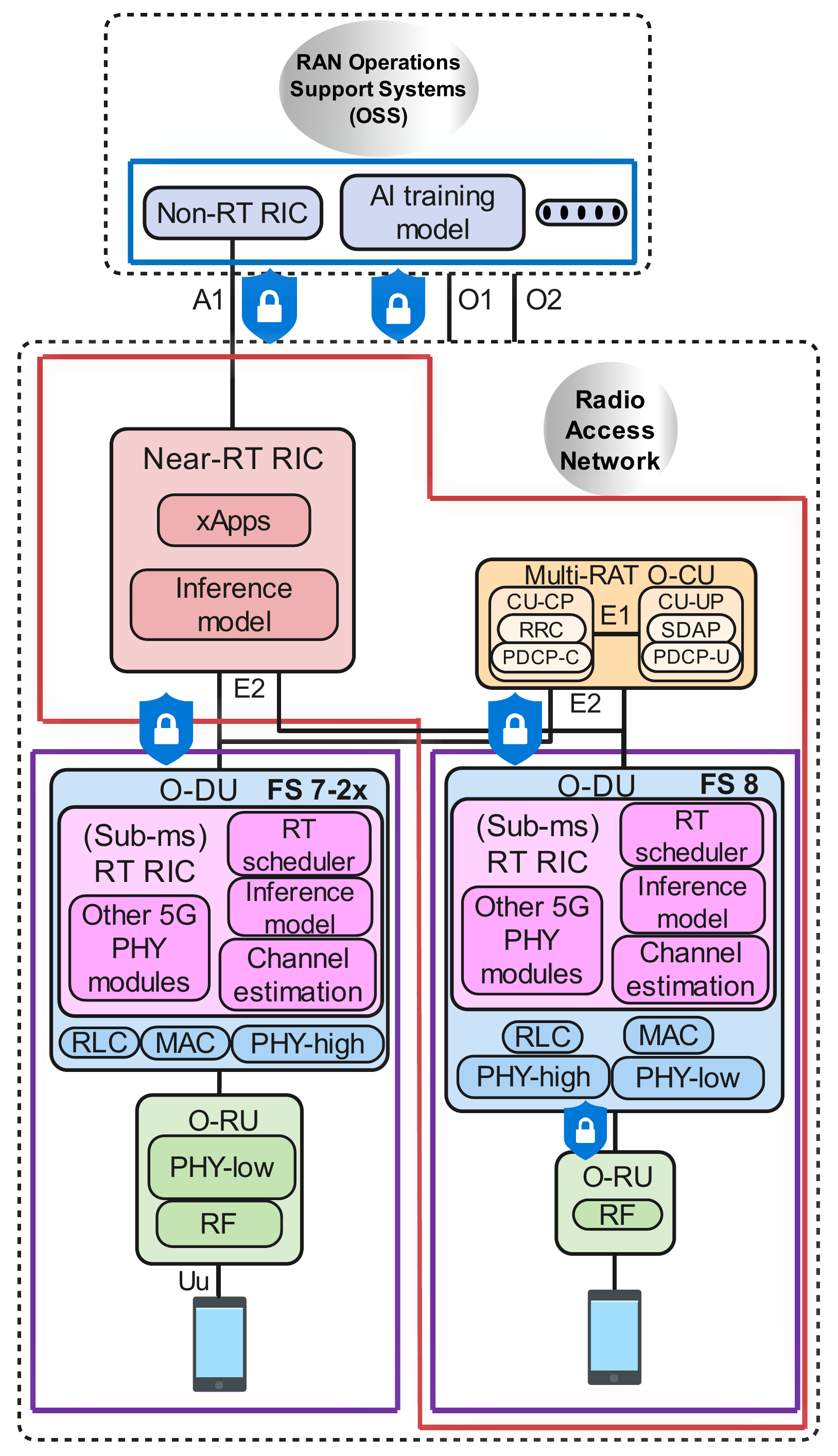}
\vspace{-7mm}
\caption{Proposed next generation O-RAN architecture extensions: O-DU/O-RU splits, RT RIC, and RT control loop. %\hl{Could you make the line spacing RT scheduler, Interence model, etc. smaller so that they fit nicely into their boxes. Channel estimation fits well, could you please adjust the other 5? If easy, put "FS 7-2x" and "FS 8" without quotes (reverse order than now).}%for next generations.
}
\label{fig:ORANNG}
\vspace{-3mm}
\end{figure}

    The O-RAN Alliance has established a security focus group %(SFG) %with its main goal being 
    which aims to define the security requirements, designs, and solutions for O-RAN security threats. %The SFG has provided possible set of solutions for the security threats, such as  
    There are more R\&D directions that need to be explored. % for improving the security and mitigate potential threats.
    These include the mutual authentication for verifying access to O-RAN system 
    and prevent malicious applications and components, mechanisms enabling trusted implementation of xApps and AI models, and % must be ensured over the O-RAN components and interfaces. 
    cryptography with secure key management, %for O-RAN through 
    including key generation, storage, rotation, and revocation. % should be integrated within the O-RAN system.
    Figure~\ref{fig:ORANNG} indicates the need for employing mutual authentication mechanisms over the O-RAN interfaces to preserve the privacy and confidentiality of the data. \textcolor{black}{%At the implementation level, the use of open-source software, including operating systems and virtual infrastructure managers increases the security risks. 
    The decoupling of hardware and software requires a trust chain from bottom up. One promising solution to mitigate security risks in this context is the adoption of a zero-trust security framework %~\cite{} 
    that assumes no implicit trust and continuously evaluates the risks.} %\hl{Should we mention a disadvantage of this architecture?}   
    %\hl{Expand here ways to go about these limitations, to remove the limitation and add new features, and what our recommendations are for R&D}
    % \textcolor{blue}{Research and development directions to explore:
    % (1) Secure interfaces, (2) Secure data against attacks, such as poisonon attacks.,
    % (3) Validate and test xApps
    % Modern approaches to improve security is the use of certificates
    % and use of Public Key Infrastructure (PKI), quantum communications for sharing the keys....}
    
    %\Vijay{I think it is important to mention about O-RAN STG group and the current approaches in consideration and potential issues with them.}

%\subsection{(Sub-millisecond) Real-time Timescale Support} 
\subsection{Deterministic Latency}
%\textcolor{blue}{Talk about the time granularity and how it cannot handle PHY layer control. Include Subsection D here, etc. (two sub-themes + latency as Vijay suggested}

%\textcolor{blue}{Latency:}
Open interfaces and multi-vendor support spur innovation. This, however, makes it more difficult to control and optimize the data and control plane latency. The general concept and latency model of the O-RAN architecture is based on the eCPRI reference model for delay management. 
    %(Vuk: I disagree with this definition. This is just the propagation delay and is negligible) The transmission delay is the allocated time from when a bit leaves the transmitter until it is received at the receiver. 
    %However, these delays may not be fixed due to switching delays, thus, a range with upper and lower bounds must be considered. 
    %For an efficient management of latency requirements, 
    
    The O-RU constraint is typically preset as it depends on the hardware. %, whereas and the control of the transport is not possible. 
    The signal transfer between the O-RU and O-DU is expected to have a relative time error over the synchronization plane %(S-plane) %within a limit 
    of 3~$\mu$s ($\pm$ 1.5~$\mu$s). 
    In order to ensure the reception of data %at the receiver 
    within the established delay boundaries, the transmitter and the receiver need to define a %proper
    transmission and a reception window, respectively.  %the transmission and reception 
    These windows are defined with respect to each other, where the reception window must be greater than or equal to the sum of the transmission window and the transport variations for the uplink and the downlink transmissions. 
    %Thus, it is possible to define the latency constraints for any one of O-DU, transport, and O-RU based on knowledge of the other two. However,
    Based on the O-RAN specifications, the parameters used for defining the reception and transmission windows must be reported with an accuracy of 200~ns. %\textcolor{purple}{[REF]}. 

    The above timing requirements %that are specified by the O-RAN 
    cannot be %fully
    guaranteed %, in particular 
    when an O-RAN system communicates with another system that does not %imply or
    follow the O-RAN delay specifications. %More specifically,
    This may lead to cases where the packets are transmitted or received too early, before the start of the window, or too late, after the end of the window. Such packets should be discarded. %and never processed, 
    However, %due to the lack of this specifications in the O-RAN, it is possible 
    the specification do not impede processing too early or too late %or too early
    packets which may disturb the O-RAN timing process of control or user plane packets. 
    
    When we compare this to the CPRI and eCPRI timing specifications, we notice that the CPRI uses strictly periodic scheduling with the k28.5 time marker that makes the latency and delay measurement more deterministic. The eCPRI, on the other hand, relies on statistical synchronization without a fixed time marker such as the precision time protocol.
    
    {\textcolor{black}{A future R\&D direction to tackle this challenge is the dynamic FS}~\cite{Pamuklu_2021}. The dynamic FS relies on varying the placement of the different network functions of the O-RAN architecture to satisfy the latency requirements based on %varying 
    feedback from the network. Figure~\ref{fig:ORANNG} illustrates how the future O-RAN architecture can support multiple FSs supported by a single near-RT RIC to %be able to support 
    satisfy different timing requirements.} %can take the full advantage of the disaggregated and virtualized RAN by employing the dynamic splits where network functions are placed   }
    % (PTP). %This has implications on the latency. 
    % Vuk: unless we say what implications it has, I think we should not simply say that it has implications. 
    
    %\hl{Expand here ways to go about these limitations, to remove the limitation and add new features, and what our recommendations are for R\&D}

\subsection{PHY Layer RT Control}
\textcolor{black}{As discussed previously, one of the foremost benefits of the near-RT RIC is the intelligent and fine-grained RAN control functionalities at both the O-CU and the O-DU through the E2 interface. 
%%Depending on the FS, almost the entire RAN protocol stack is exposed to the near-RT RIC.
%\textcolor{brown}{Though these are major advances towards some functionalities which haven't been considered by the O-RAN Alliance due to current technological limitations. 
%%However, the O-RAN Alliance does not explore the RF and PHY layer characteristics %of the RAN 
%%and relies on the 3GPP specifications. % for functionalities, such as resource mapping, modulation, and coding.
%%In other words, it is assumed that traditional signal processing algorithms/technologies implement the RF, intermediate frequency, and baseband processing procedures. 
%This hinders the realization of the full potential of the O-RAN paradigm. For example, the 
While the near-RT RIC could perform rudimentary control of PHY layer functions, such as reducing the beam search space, %Additionally, the timescale at which the near-RT RIC operates.
}
\textcolor{black}{%This restricts the near-RT RIC in unleashing its full potential since it is able to control rudimentary functionalities at the PHY layer.
%On the other hand, 
the timescale at which it operates (10 ms -- 1 s) makes it difficult to control many of the PHY layer %adjustments %characteristics
processes. % which undergo dynamic changes at sub-millisecond granularity. 
\textcolor{black}{This necessitates the introduction of a (sub-millisecond) real-time controller hosted at either the O-CU or the O-DU}. % to perform actions at the PHY layer.
%%To further complicate matters, the O-RUs and UEs are not standardized %of this makes it more difficult. RU/UE
%%and standardizing them is not an easy process. 
}

%We propose a new entity for O-RAN: the RT RIC (Fig.~\ref{fig:ORANNG}). 

{\color{black}{We propose integrating a third control loop, %termed, (sub-millisecond)
the real-time control loop enabled by the RT RIC, which can possibly be situated in the O-CU, O-DU, or O-RU.}} Although the RT RIC can host any type of lower layer RAN control functionality, which we call \textit{zApps}---\textcolor{black}{%The \textit{zApps} are 
third-party control applications hosted in the envisioned RT RIC---}%\footnote{\textcolor{blue}{We introduce the term %the terminology 
%\textit{zApp} to refer to third-party control applications hosted in the envisioned RT RIC.%, and it does not have any bearing with 3GPP or O-RAN standard specifications.
%}}
the introduction of AI will enable dealing with the complexity of the PHY layer, heterogeneous resources, and operating environments for building %flexible
\textcolor{black}{highly configurable, yet manageable} RANs. Research has shown that the performance of the RAN can be improved by employing data-driven learning techniques\textcolor{black}{~\cite{Pamuklu_2021, polese2021colo, VTC-2021ai, RTRIC}.} %\hl{multiple citations R1C1}. %For example, RF non-linearity, which is a transmitter or receiver problem, can be modeled and mitigated with a neural network as part of the baseband processing~\cite{VTC-2021ai}. 
Obstacles for realizing this include the compute power of the nodes (O-CU/O-DU), the energy efficiency, the ability of AI models to deliver decisions at sub-millisecond timescale (response time constraints), and the amount of signaling overhead for low-level PHY layer control. These are critical factors that might have an adverse impact on fulfilling the promises of low-latency next generation cellular networks. %, specifically ultra-reliable low latency communications. % case in point being URLLC communications.

\textcolor{black}{Promising solutions to the above problem include running the %se algorithms \hl{(zApps?)} 
zApps at near-constant time complexity combined with subject-matter expert validation~\cite{RTRIC} and developing a new class of highly accurate and lightweight AI algorithms, such as echo state networks, which are less data intensive. % to enable the new RT control loop and the RT RIC. % that we envisage for next generation O-RAN. 
The integration of RAN hardware acceleration %architectures
within O-RAN %~\cite{} 
is expected to augment the performance of x86 %and ARM 
based server platforms %which will be 
that implement O-CUs and O-DUs.} 

%\hl{Do we need to provide any points on how the proposed RT-RIC addition to the architecture overcomes the challenges we outlined?} 
%\hl{Expand here ways to go about these limitations, to remove the limitation and add new features, and what our recommendations are for R\&D}

%real time control loop - 

   % \hl{Expand here ways to go about these limitations, to remove the limitation and add new features, and what our recommendations are for R\&D}
%\end{itemize}

\subsection{\textcolor{black}{Testing of AI-Driven O-RAN Functionalities}}
%\textcolor{blue}{The interfaces defined Additional RAN functionalities may need additional interfaces from the CN to the RIC, e.g. for age of information (AoI)-based network control.
%Support different data types.}

%\subsection{Testing Framework}
%\textcolor{blue}{Support testing and verification of xApps through well-defined interfaces....}

%\textcolor{blue}{\textbf{Testing:} 
\textcolor{black}{Wireless network intelligence is enabled in O-RAN through AI-driven xApps and deployments in the near-RT and non-RT RICs. However, %of the O-RAN architecture is not guaranteed because of 
the unpredictable behavior of the AI algorithms, specifically when considering a closed-loop AI controller, may lead to unstable system configurations and performance losses. Poorly-designed AI solutions can degrade the quality of decisions made by the xApps, rApps, and the envisioned zApps. % and therefore, it can inflict the automation performance of RAN processes. 
In addition, xApps, rApps, and zApps may be incompatible and have opposing objectives that can lead to unstable network operations. %Examples include incompatible xApps or runtime data deviating from what was used for training.   
}

\textcolor{black}{It is therefore critical to evaluate and test the stability and robustness %performance 
of the AI algorithms to be deployed %, especially for large scale environments 
to avoid untested configurations, conflicting decisions, or exposure of vulnerabilities. % desirable operation, or any shortcomings or unoptimized solutions.
In addition to standardized predeployment test, measurement, and certification procedures, testing and maintenance needs to be an integral part of the O-RAN operations. Such online testing procedures within the intelligent controllers of O-RAN need to monitor and assess the employed AI algorithms for adapting to unfamiliar %measurements 
data or operational environments, such as excessive noise or RF interference, unexpected signals, or conflicting messages from the network or the users. %RF interference hunting is a challenging problem in cellular networks because the standards are developed for operation in known radio environments with known types of interference sources (e.g., inter-cell interference). %that has not been provided during the training or testing phases to enable more 
%generalization and robustness to noise, which ensure efficient data-driven solutions. 
In addition, the testing framework should be capable of defining the competence of the AI algorithms to cope with the randomness and variations of the data and cases of incomplete, uncertain, or missing training data. 
The interpretability and explainability of AI algorithms \textcolor{black}{have to} be considered within the O-RAN testing workflow where more research and data are needed. The interpretability shows how accurate the output of the model is given a certain input. The explainability illustrates how the internal mechanisms and designs of the AI model affect the %achieved 
output. 
}

%\hl{This first sentence is unclear to me}
%%The scope of testing process %within the O-RAN architecture 
Testing should encompass the AI models and the data because AI models may fail despite regular data or they may fail because false data is injected. \textcolor{black}{We recommend %considering future services and applications that can be integrated with 
integrating services into the AI workflow for such testing and validation %. Specifically, these new add-on are able 
and into next-generation RAN processes for delivering high-fidelity live representations of all functions that are running on O-RAN deployments, where one O-CU, O-DU, and O-RU configuration may differ from another within the same operator and across operators and services. }

\section{Conclusions}
\label{sec:conclusions}

\textcolor{black}{%Open Radio Access Network (O-RAN) 
O-RAN defines a %envisioned as the
disruptive architecture %technology 
for realizing next generation wireless networks. It %, due to it's 
specifies open interfaces and logical elements that support intelligence, disaggregation, softwarization, virtualization, and collaboration. We conducted a survey among researchers, developers, and practitioners to explore their interests in O-RAN %technology 
and %report 
their needs for 6G R\&D. %findings in this paper. 
%Furthermore, 
As a result of our survey, we identify %key
critical limitations of the current O-RAN specifications in terms of security, latency, real-time control, and testing of AI-based RAN control applications. 
We outline the technologies and R\&D opportunities %d key research directions and made some key R\&D recommendations which can potentially 
to overcome these limitations and extend the O-RAN architectural capabilities.}
%to realize %, and help realize the 
%wide adoption of O-RAN for future 6G wireless networks.}
% spur innovations in the O-RAN and eventually 

\section*{Acknowledgement}
This work was supported in part by the National Science Foundation under grant numbers 2120411 and 2120442.

\balance

\bibliographystyle{IEEEtran}
\typeout{}
\bibliography{main}

\section*{Biographies}
\footnotesize
\vspace{0.2cm}
\noindent
\textbf{Aly Sabri Abdalla} (asa298@msstate.edu)
is a PhD candidate in the Department of Electrical and Computer Engineering at Mississippi State University, Starkville, MS, USA. His research interests are on scheduling, congestion control and wireless security for vehicular ad-hoc and UAV networks.

\vspace{0.2cm}
\noindent
\textbf{Pratheek S. Upadhyaya} (pratheek@vt.edu)
is a PhD student in the Department of Electrical and Computer Engineering at Virginia Polytechnic and State University, Blacksburg, VA, USA. His research interests include scheduling, 5G/next-G communications, AI/ML applications in wireless communications and spectrum sharing.

\vspace{0.2cm}
\noindent
\textbf{Vijay K. Shah} (vshah22@gmu.edu) is an assistant professor of cybersecurity engineering at George Mason University, Farifax, VA, USA. His research interests include 5G/next-G communications, AI/ML, wireless security, and wireless testbed development. 
%\textbf{Keith Powell} (kp1747@msstate.edu) is pursuing a PhD degree in the Department of Electrical and Computer Engineering at Mississippi State University, Starkville, MS, USA. His research interests include software radio platforms, UAV communications, and embedded systems.

\vspace{0.2cm}
\noindent
%\textbf{Walaa AlQwider} (wq27@msstate.edu) is pursuing a PhD degree in the Department of Electrical and Computer Engineering at Mississippi State University, Starkville, MS, USA. Her research interests include spectrum sharing between active and passive systems and AI-enhanced communications.

\vspace{0.2cm}
\noindent
\textbf{Vuk Marojevic} (vuk.marojevic@msstate.edu) is an associate professor in electrical and computer engineering at Mississippi State University, Starkville, MS, USA. His research interests include resource management, vehicle-to-everything communications and wireless security with application to cellular communications, mission-critical networks, and unmanned aircraft systems.

\end{document}